\documentclass[%
 reprint,
superscriptaddress,
 amsmath,amssymb,
 aps,
 longbibliography
]{revtex4-1}

\usepackage{mathtools}

\usepackage[usenames]{color}
\usepackage{amssymb}
\usepackage{amsmath} 

\usepackage{esdiff}
\usepackage{amsthm}
\usepackage{placeins}
\usepackage{makeidx}
\usepackage{url}
\usepackage{float}
\usepackage[caption = false]{subfig}
\usepackage[normalem]{ulem}
\usepackage{tabto}

\usepackage{graphicx}
\usepackage{dcolumn}
\usepackage{bm}

\begin{document}

\title{\textbf{Growth strategy determines aspects of brain performance}}
 
\author{Ana P. Mill\'an$^{a*}$, J.J. Torres$^{a}$, S. Johnson$^{b}$ and
J. Marro$^{a}$\\
{\small{}$^{a}$ Institute }\textit{\small{}Carlos I}{\small{} for
Theoretical and Computational Physics,}\\
{\small{}University of Granada, Spain}\\
{\small{}$^{b}$ School of Mathematics, University of Birmingham,} \\
{\small{} Edgbaston B15 2TT, UK.} \\
{\small{}$*$ Corresponding author (apmillan@ugr.es).}}

\begin{abstract}
\textbf{The interplay between structure and function is crucial in determining some emerging properties of many natural systems. 
Here we use an adaptive neural network model  
inspired in observations of synaptic pruning that couples activity and topological  
dynamics and reproduces experimental temporal profiles of synaptic density, including an initial transient period of relatively high synaptic connectivity.
Using a simplified framework, we prove that the existence of this transient is critical in providing ordered stationary states that have the property of being able to store stable memories. In fact, there is a discontinuous phase transition between the ordered memory phase and a disordered one as a function of the initial transient synaptic density. 
We also show that intermediate synaptic density values are optimal in order to obtain these stable memory states with a minimum energy consumption, and that ultimately it is the transient heterogeneity in the network what determines the stationary state.
Our results here could explain why the pruning curves observed in actual brain areas present their characteristic temporal profiles and, eventually, anomalies such as autism and schizophrenia associated, respectively, with a deficit or an excess of pruning.}

 \end{abstract}

\maketitle


Complex networks are ubiquitous in nature: almost every biological and social system as well as many man-made structures develop intricate relations among its components, resulting in a network configuration that is usually far from being homogeneous \cite{boccaletti2006complex,newman2010networks}.
Research on complex networks has received a tremendous amount of attention over recent decades, in order both to understand and protect natural networks and to optimize technical designs.  
Most studied networks have thus been shown to present non-trivial topological features, such as high clustering and short minimum paths (small-worldness), modular structure, and cost-efficient wiring \cite{eguiluz2005scale,reka2005scalefree,gastner2015topology}.
A main global observation following from these studies is that the majority of known networks exhibit also highly heterogeneous degree distributions (where the degree of a node is its number of neighbors) and, except in the case of social networks, negative degree-degree correlations -- a property known as disassortativity \cite{odor2013}.
In other words, many networks of interest include a small number of highly connected nodes, called hubs, which tend to be connected to low-degree nodes \cite{newman2003structure}.
Interesting enough, these properties can influence the emerging dynamics of complex systems. 
For instance, it is known that both degree heterogeneity and degree-degree correlations strongly influence the signal to noise ratio in certain dynamical systems \cite{maslov2002specificity,torres2004influence,franciscis2011enhancing} and that the synchronization properties of a complex network strongly depend on its dimension \cite{eytan2006dynamics,yoCNM}.

In order to understand how such non-trivial networked structures come about, much work has gone into investigating mechanisms of network evolution. Models in which  networks are gradually formed, for instance by addition and/or deletion of nodes and edges, or by the rewiring of the latter, have been studied in various contexts \cite{berg2004structure,johnson2009nonlinear,navlakha2015decreasing}. 
Typically, the probabilistic addition or deletion of elements at each time step is a function of the   structure existing at that time. 
For example, in the familiar Barab{\'a}si-Albert model \cite{barabasi1999emergence}, a node's probability of receiving a new edge at a given time is proportional to its current degree. 
Certain evolution rules have thus been shown to generate network topologies with particular properties, such as small-world, scale-free, or hierarchico-modular structures \cite{watts1998collective,barabasi1999emergence,bianconi20016network}. 
These rules often give rise to phase transitions, such that qualitatively different kinds of network topologies can ensue depending on parameters. 
The idea behind all such evolving network models is that there should be general, relatively simple ``microscopic'' mechanisms which can give rise to these complex structures without the need for previous high levels of information or a propos tuning \cite{newman2003structure,reka2005scalefree}.

In most studied networks the evolution of the topology is invariably linked to the state of the network and vice versa \cite{Gross08},
which has given rise to a novel field of study: \textit{adaptive networks} \cite{Sayama20131645}. Such a coupling between the activity within the network and the evolution of its topollogical structure creates a feedback loop between dynamics and topology. In this way some interesting dynamic phenomena occur repeatedly in adaptive networks, such as the formation of complex topologies, robust dynamical self-organization, spontaneous emergence of different classes of nodes, generally through a  complex mutual dynamics of both activity and topology \cite{vazquez08,su13,wiedermann15}.

This framework can be naturally applied to try to understand brain development. 
The mammalian brain is initially formed through an initial rapid proliferation of synapses.
Synaptic density thus reaches a peak during early infancy and from then on it begins a steady decline down to about half this value later in life, in a process known as \textit{synaptic pruning} \cite{chechik1998synaptic,iglesias2005dynamics}.
It is believed that the reason for reducing synaptic density is becoming more energetically efficient \cite{chechik1999neuronal}. 
But then a question arises, why not begin life with the optimal synaptic density?
Recent studies have suggested that details of synaptic pruning may have large implications on high-level brain functions, and they have been related to the emergence of some neurological disorders such as autism and schizophrenia \cite{faludi2011synaptic,autismo1}.
However, little is so far known about the influence of synaptic pruning and the relatively high synaptic connectivity during early brain development on its performance. 

Here we consider a computational model of synaptic pruning that includes an initial transient period of high synaptic density, and may thus serve to analyze what factors affect 
both pruning efficacy and details of the stationary state of the system. 
Our model adatps a biologically inspired evolving neural network model  \cite{yo2017,FCN} that  couples an attractor neural network \cite{amit1992modeling} with a preferential attachment network evolution \cite{johnson2010evolving}, based on the fact that synaptic growth and death are related to neural activity \cite{klintsova1999synaptic}. 
This setup creates a feedback loop between structure and dynamics, leading to two qualitatively different kinds of behaviors, as it has been previously shown \cite{yo2017,FCN}. 
In one of them, the network structure becomes heterogeneous and dissasortative and the system then displays good memory performance. In the other, the structure remains homogeneous and incapable of pattern retrieval.
We also show that the inclusion of an initial, transient rapid growth, even in a simple case, introduces a dependence on the initial synaptic density that increases network performance, in terms of memory retrieval, even in the presence of high levels of noise.
The basic mechanism which illustrates our model here needs not be restricted to neural networks, but may help understanding also how other structures form, e. g., in the case of
protein interaction networks which also change in evolutionary time scales in a way that is related to their physiological activity \cite{berg2004structure}. 
In fact, as it is the case of our model, most biological networks change with time so that pruning may be a general mechanism for network optimization trying to minimize energy consumption in an environment of limited resources without the need for a great amount of information specifying the initial topology.


\subsection*{Synaptic pruning model}
The synaptic pruning model in this work couples a traditional associative memory model, the Amari-Hopfield model \cite{amit1992modeling}, with a preferential attachment model for network evolution \cite{johnson2010evolving,yo2017,FCN}. 
The system consists of an undirected $N$-node network whose nodes may be interpreted as neurons, whereas edges stand for synapses. 
Network structure is defined by the adjacency matrix $e_{ij}=\lbrace 1,0 \rbrace$, indicating the presence or absence of an edge, respectively. 
Though the model may be easily generalized in this and other details, each neuron $i$ is modeled by a stochastic binary variable $s_i(t)=\{0,1\}$, indicating a silent or firing neuron. 
$s_i(t)$ follows an Amari-Hopfield dynamics where the level of stochasticity is characterized by a noise parameter or temperature $T$, where $T=0$ corresponds to the deterministic limit \cite{amit1992modeling} (see \ref{methods}). 
Each existing synapse is characterized by its synaptic weight $w_{ij}$, which is a real variable defined according to a Hebbian learning rule so that the pattern of activity  $\left\lbrace \xi_i = 0,1 \right\rbrace$ is made to be an attractor of the dynamics, therefore called a ``memory'' \cite{amit1992modeling}.
The macroscopic state of the system may be characterized by the overlap of the network with the memorized pattern, namely
\begin{equation}\label{eq:ov}
m(t) = \frac{1}{Na_0(1-a_0)} \sum_{i=1}^N (\xi_i-a_0)s_i
\end{equation}
where $a_0 = \left\langle \xi_i \right\rangle$ is the mean activation of the pattern, introduced for normalization purposes. $m(t)=1$ indicates that the state of the network is equal to the pattern, and the system is in a memory state, whereas $m(t)=0$ corresponds to the non-memory state. 
In a fully connected network, the stationary value $\bar{m} \equiv m(t\to \infty)$ undergoes a continuous transition from a phase of memory recovery ($\bar{m}\rightarrow 1$) to one dominated by noise ($\bar{m} \approx 0$) as a function of $T$, at the critical value $T_c=1$.

The topology changes in time following a Markov process given by the probabilities $\widetilde{P}_i^g$ and $\widetilde{P}_i^l$ that each node $i$ has to gain and loose an edge at time $t$. These are assumed to factorize in two terms, $\widetilde{P}_i^g=u(\kappa)\widetilde{\pi}(I_i)$, $\widetilde{P}_i^l = d(\kappa)\widetilde{\eta}(I_i)$, where $\kappa$ is the mean connectivity or degree of the nodes in the network and $I_i$ stands for the incoming current at each neuron from its neighbors. 
Here $u$ and $d$ represent a global dependence to account for the fact that these processes rely in some way on the diffusion of different molecules through large areas of tissue, for which $\kappa$ is taken as a proxy \citep{klintsova1999synaptic}.
Provided that $\widetilde{\pi}(I_i)$ and $\widetilde{\eta}(I_i)$ are normalized over the network, the evolution of $\kappa(t)$ only depends on $u(\kappa)$ and $d(\kappa)$, and it is given  \cite{johnson2010evolving,yo2017,FCN} by
\begin{equation}
\frac{d\kappa(t)}{dt} = 2 \left[ u\left(\kappa(t)\right) - d\left( \kappa(t)\right) \right].
\end{equation} 
Experimental evidence indicates a fast growth of the synaptic density following birth and preceding synaptic pruning, whose impact on brain development is yet to be fully clarified \cite{navlakha2015decreasing,autismo1}.
These profiles of synaptic density during infancy 
have been reproduced before with this model  \cite{yo2017,FCN} by considering an initial, fast growth of synapses followed by synaptic pruning, namely 
\begin{equation}\label{eq:uyd}
\begin{array}{lll}
u\left(\kappa\right) &=& \frac{n}{N}\left(1-\frac{\kappa}{2\kappa_{\infty}} + a_g\exp(-t/\tau_g) \right) \\
d\left(\kappa\right) &=& \frac{n}{N}\frac{\kappa}{2\kappa_{\infty}},
\end{array}
\end{equation}
where $\tau_g$ and $a_g$ control the time-scale and intensity of the initial growth, $\kappa_\infty$ is the stationary mean connectivity, and $n/N$ characterizes the speed of synaptic growth and death. This leads to
$
\frac{d\kappa}{dt} = \frac{2n}{N}\left[ 1 - \frac{\kappa(t)}{\kappa_\infty} + a_g\exp(-t/\tau_g) \right],
$
which has the solution
\begin{equation}\label{eq:ktg}
\kappa(t) = \kappa_\infty \left[ 1 -b\exp(-t/\tau_g) +c\exp(-t/\tau_p) \right],
\end{equation}
where $\tau_p=N\kappa_\infty/2n$ sets the temporal scale of the pruning process,  $b=a_g\tau_g\left(\tau_p-\tau_g\right)^{-1}$ and $c= \kappa_0/\kappa_\infty +b -1$. This model for $\kappa(t)$ (dashed black line in figure $\ref{fig:fig1}a$) has been shown to reproduce experimental data of the mean synaptic density in the cortex in humans and rodents \cite{yo2017}.

The local probabilities $\widetilde{\pi}$ and $\widetilde{\eta}$ introduce a dependence on the physiological state of the neurons, and account for local heterogeneity in the network. Following previous studies, we take $\widetilde{\pi}(I_i) \propto I_i ^ \alpha$ ($\alpha>0$), $\widetilde{\eta}(I_i) \propto I_i$, which corresponds to synapses being chosen at random for removal, which can be seen as a first order approximation of pruning dynamics \cite{johnson2009nonlinear}.
Network structure is characterized by the homogeneity parameter $g(t)$,
\begin{equation}
g(t) = \exp \left(-\sigma_k^2(t)/\kappa(t)\right),
\end{equation} 
where $\sigma_k^2(t)$ is the variance of the degrees of the nodes. $g(t)$ equals $1$ if $p(k)=\delta_{k_0,k}$ (homogeneous network) and tends to $0$ for highly heterogeneous (bimodal) networks. 
In a topological limit of this model ($I_i \rightarrow k_i$), the stationary value $\bar{g}\equiv g(t\to\infty)$ undergoes a continuous phase transition from homogeneous networks to heterogeneous ones at $\alpha_c=1$ \cite{johnson2010evolving}.


Previous work have analyzed the effect of the coupling between the local structure of the network and the physiological dynamics \cite{yo2017,FCN} during synaptic pruning ($a=0$ in eq. $(\ref{eq:uyd})$). 
Depending on $\alpha$, the parameter controlling the heterogeneity of the pruning dynamics, and the temperature $T$ setting the thermal noise system, three phases were shown to appear: a homogeneous memory phase when both $\alpha$ and $T$ are low, in which the network displays memory and its structure is homogeneous; a heterogeneous memory phase for high $\alpha$, in which the network is bimodal (appearance of hubs); and a homogeneous noisy phase for high noise $T$. 
Finally, a bistability region was also shown to appear between the heterogeneous memory and the homogeneous noisy phases (which corresponds to moderate $\alpha$ values, $1<\alpha < 2$, and high temperature, $T>1$) as a consequence of the coupling introduced by the model. 
This is because heterogeneous networks are more tolerant to thermal noise than homogeneous ones, so that the critical temperature $T_c$ separating the memory and non-memory phases diverges from $T_c=1$ for homogeneous networks to $T_c\to \infty$ for heterogeneous ones, where the presence of \textit{hubs}  stabilizes the memories \cite{torres2004influence,franciscis2011enhancing}. 
In this region the stationary state of the system depends on its initial configuration: networks that are initially heterogeneous display memory and enhanced heterogeneity during the whole evolution of the system, whereas homogeneous ones fall into the noisy state. 
Consequently, the initial overgrowth of synapses could have important consequences in the emerging behavior of the system in this region, since it changes the state of the system when synaptic pruning begins. Therefore, in the following we analyze the effect of the non-trivial transient of high connectivity preceding synaptic pruning on the dynamics of the system  on the bistability region (in particular, we choose $\alpha = 1.2$, $T=1.3$).

\subsection*{Linear approximation}

\begin{figure}
\begin{centering}
\includegraphics[width=1.0\columnwidth]{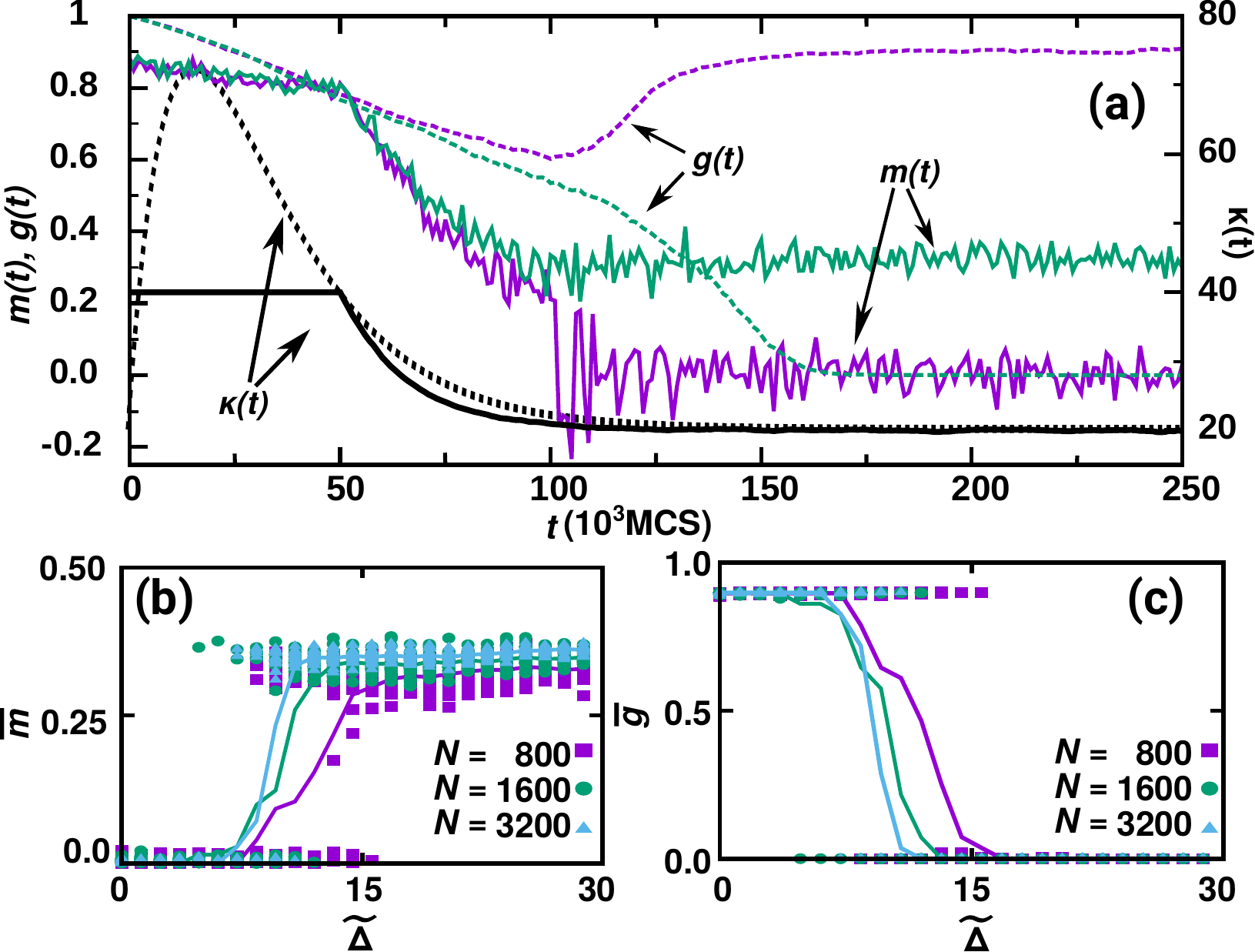}
\par\end{centering} 
\caption{\textbf{(a)} The black lines represent $\kappa(t)$ as given by the realistic model (black dashed line) and the linear approximation model (black solid line). The colored lines show two exemplary temporal evolutions of the linear model with $\kappa(t)$ as given by the black solid line. 
We represent $m(t)$ (colored solid lines) and $g(t)$ (colored dashed lines) for two realizations leading to two qualitatively different steady states, implying multistability in the system. 
In green, a series in which the network keeps memory ($\bar{m}=0.35$) and heterogeneity ($\bar{g}=1.0$) in the steady state. 
In purple, the opposite example. All parameters are the same in both situations ($N=1600$, $n=2$, $\kappa_0=40$ and $\kappa_\infty = 20$ and $\Delta = 5\cdot 10 ^4 MCS$). 
\textbf{(b)} $\bar{m}(\widetilde{\Delta})$ and \textbf{(c)} $\bar{g}(\widetilde{\Delta})$ curves for different system sizes, where $\widetilde{\Delta}=\Delta/\tau_p$.
Results are for $\kappa_0=40$ and $\kappa_\infty = 20$. The parameter $n$ is scaled with the network size, so that $n = 5, 10, 20$ respectively for $N=800, 1600, 3200$. \label{fig:fig1}  }
\end{figure}

One can easily see that $\kappa(t)$ as given by eq. $(\ref{eq:ktg})$ presents a maximum $\kappa^*=\kappa(t^*)$, where $t^* = \frac{\tau_g\tau_p}{\tau_p-\tau_g} \ln \left(\frac{\tau_p b}{\tau_gc}\right)$.
Both $t^*$ and $\kappa^*$ depend non-trivially on $a$, $\tau_g$ and $\tau_p$. Therefore, in order to study the effect of the non-trivial transient of high connectivity on the emergent state of the system, here we consider a first order approximation to this realistic pruning profile.
In particular, $\kappa(t)$ is kept constant (and high), $\kappa(t)=\kappa_0$, at the onset of the evolution (see solid black line in figure $\ref{fig:fig1}a$), during a ``frozen-density'' time $\Delta$, by imposing that the same number of edges are created and destroyed, namely
\begin{equation}
u(t)=d(t)=d_0\ \forall t<\Delta,
\end{equation}
 where $d_0 = const$, and in particular we consider here $d_0 = n/N$.
Thereafter, the mean degree is allowed to vary following the pruning dynamics. This allows us to easily control the width ($\Delta$) and height ($\kappa^*= \max(\kappa_t)$) of the pruning process, which fully characterize it. 

An exemplary evolution of the coupled system is shown in figure $\ref{fig:fig1}a$. 
The initial ``frozen-density'' period ($t<\Delta$), provides a non-trivial transient of network evolution during which edges are added and removed but $\kappa(t)$ is kept constant and equal to $\kappa_0$. 
If $\kappa_0$ is sufficiently large, the system can perform memory retrieval throughout this period even though $T>1$, as indicated by an overlap $m(t)$ significantly different from zero. 
Given that the topological dynamics is also taking place, there is meanwhile an underlying rewiring process that starts creating hubs and heterogeneity (given that $\alpha>1$) and thus $g(t)$ decreases.
Therefore, due to the coupling between memory and heterogeneity, if the network maintains memory it continues to heterogenize, creating a feedback between neural dynamics and topology.
Once $t>\Delta$ synaptic pruning begins, and the system can either fall into the noise state and lose its heterogeneity (purple lines in figure $\ref{fig:fig1}a$), or remain in the heterogeneous memory state (green lines), showing multistability.
If the system remains in the memory state, it continues to heterogenize ($g(t)\rightarrow0$), to the point that it can maintain memory performance  as $\kappa(t) \rightarrow\kappa_\infty$ (dashed green line in figure $\ref{fig:fig1}a$). 
On the other hand, if the neural network falls into the noisy state ($m(t) \approx 0$), neural activity -- and hence synaptic growth and death -- becomes uncorrelated with node degree, and the topology reverts gradually to a more homogeneous configuration ($g(t)\rightarrow 1$, dashed purple line), incapable of memory (due to the high noise).

In particular, for a given $\kappa_0$ there is a discontinuous phase transition as $\Delta$ is increased from a phase governed by noise, $\bar{m}\to0$, in which networks are homogeneous, $\bar{g}\to 1$ (which we shall call homogeneous noisy phase) to one where networks are heterogeneous, $\bar{g}\to 0$, and they display memory, $\bar{m}>0$ (therefore called heterogeneous memory phase), as shown respectively in panels $(b)$ and $(c)$ of figure $\ref{fig:fig1}$ (where $\Delta$ has been re-scaled as $\widetilde{\Delta}=\Delta/\tau_p$).
A finite size analysis shows that the results hold for increasing network size (figure $\ref{fig:fig1}b,c$ is for $N=800,\ 1600$ and $3200$). 


\vspace{5mm}

\subsection*{Non-linear effect of the initial density}
\begin{figure}
\centering
\includegraphics[width=1.0\columnwidth]{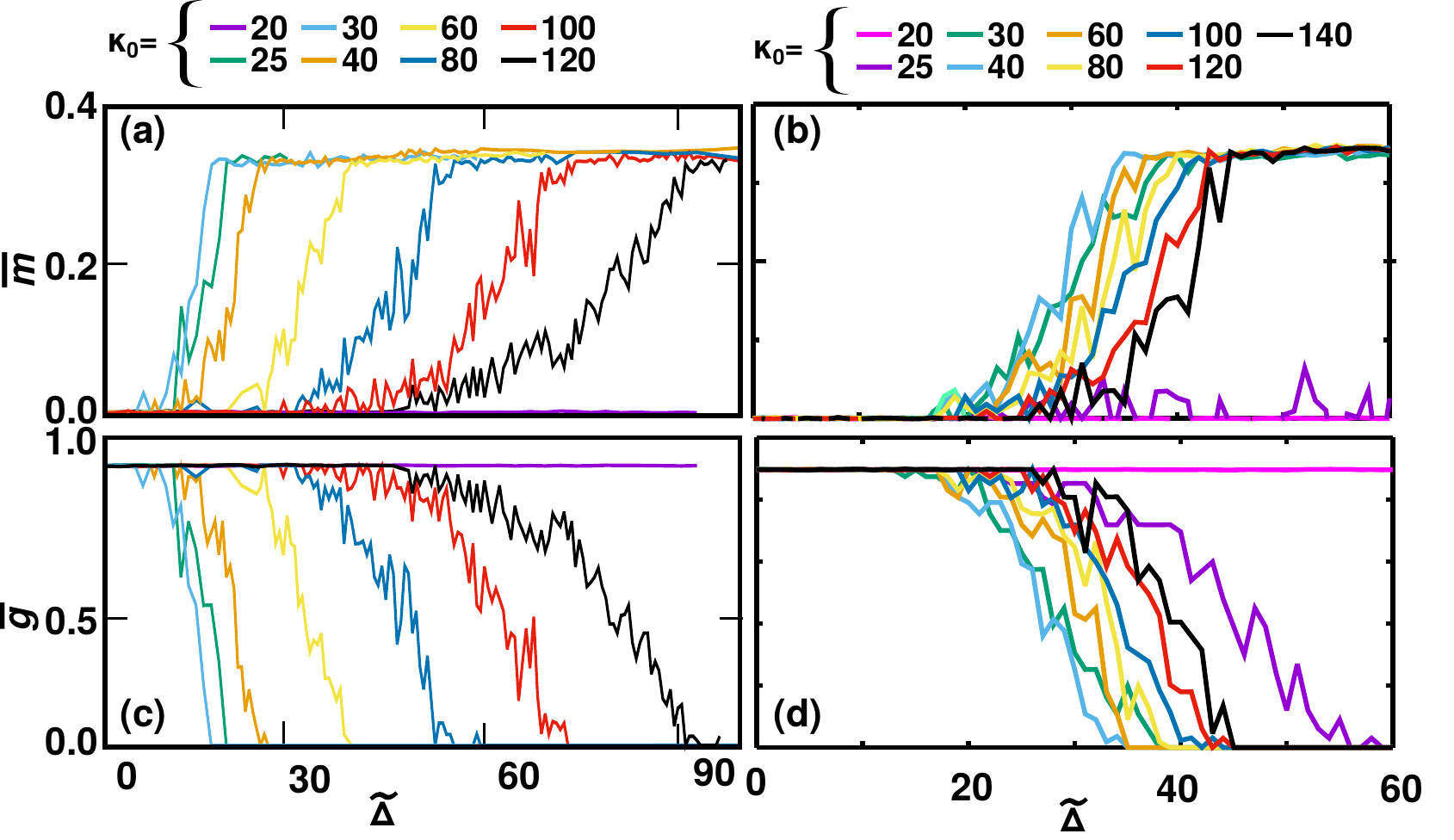}
\caption{Stationary mean values of the overlap, $\bar{m}$ (panels $(a)$ and $(b)$) and homogeneity, $\bar{g}$ (panels $(c)$ and $(d)$) respectively for models A (panels $(a)$ and $(c)$) and B (panels $(b)$ and $(d)$). Results are shown as function of $\widetilde{\Delta}$ and for different values of $\kappa_0$ as indicated by the labels on top of the diagrams.
$N=1600$ and other parameter values as in figure $\ref{fig:fig1}$. 
\label{fig:fig2}}
\end{figure}

The initial density $\kappa_0$ also has a major effect on the dynamics, determining whether the system will be able to maintain memory retrieval and the minimum $\widetilde{\Delta}$ necessary for it (see figure $\ref{fig:fig2}$).

In the case $\kappa_{0}=\kappa_{\infty}$, $\kappa(t)$ is trivially constant and, given that $T>1$, the system falls into the noisy state regardless of $\widetilde{\Delta}$ (see $\kappa_0 = 20$ in figure $\ref{fig:fig2}a,c$). This continues up to slightly higher initial densities (up to $\kappa_{0}=25$), where the memory state is reached even for very low $\widetilde{\Delta}$. 
One might expect that the memory state would become easier to reach with higher $\kappa_0$ but, in fact, the opposite effect is obtained,
and networks with increasing $\kappa_0$ take longer $\widetilde{\Delta}$ to reach the heterogeneous memory phase,  
for $\kappa_0\gg \kappa_\infty$.
This apparent paradox is explained after a deeper look at the system. 
In fact, large $\kappa_0$ implies that networks are initially more homogeneous and take more time to become heterogeneous under the topology dynamics. 
Besides, more edges have to be pruned to make a significant change in the network, which slows down network evolution.
In consequence, 
very highly connected networks are more likely to fall into the noise state for a given $\widetilde{\Delta}$. 

\begin{figure}
\begin{centering}
\includegraphics[width=1.0\columnwidth]{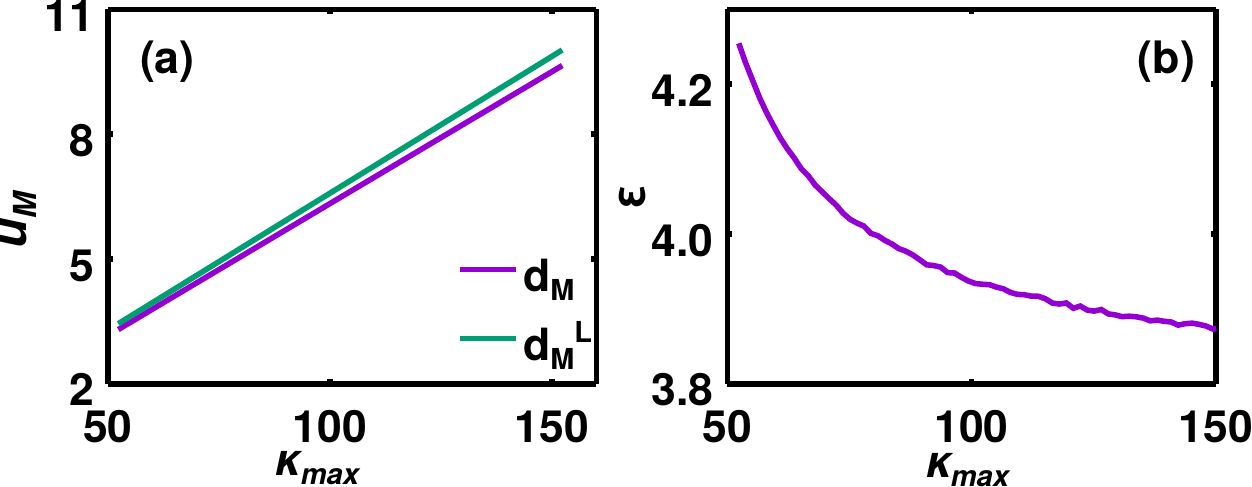}
\par\end{centering} 
\caption{\textbf{(a)} Mean proportion of edges removed between $t_1$ and $t_2$ in the realistic model ($d_m$, purple line) and the linear approximation ($d_m^L$, green line). \textbf{(b)} Normalized error associated with the linear approximation $\varepsilon$. Parameters as in figure $\ref{fig:fig2}$.  \label{fig:fig2B}  }
\end{figure}

One relevant question at this point is how this result depends on the particular definition of $d_0$, 
since so far we have considered $d_0=const\ \forall \kappa_0,\Delta$. However, one may argue that the density of synaptic turnover should depend on the number of existing synapses, $d_0=f(\kappa_0)$.
In a more real scenario, one can also define $d_0$ so as to approximate the mean number $d_m$ of links added and removed in the realistic model  between the two times $t_1$ and $t_2$ such that $\kappa(t_1) =\kappa(t_2)=\kappa_0$, $t_1<t_2$, namely $d_m \equiv 1/(t_2-t_1) \sum_{t=t_1}^{t_2} d(\kappa(t))$.
A parabolic approximation of $\kappa(t)$ as given by eq. $(\ref{eq:ktg})$ 
around $\kappa^*$ gives for $d_m$
\begin{equation} \label{eq:dm}
d_m \approx d_m^L \equiv \frac{1}{4\tau_p} \left(\kappa^* + \frac{1}{6}D\Delta_\kappa^2\kappa_\infty \right),
\end{equation}
where $D=\tau_g^{-2}\left(\kappa^*/\kappa_\infty -1 \right) - c\left(\tau_g^{-2}-\tau_p^{-2} \right)e^{-t^*/\tau_p}$ and $\Delta_\kappa = \kappa^*-\kappa_0$. 
As shown in figure $\ref{fig:fig2B}$, 
the percentage of error associated with the linear approximation ($\varepsilon \equiv 100 \times |d_m - d_m^L|/d_m$) is small. 
This suggest a scaling relation $d_0\propto \kappa_0$, and in particular one can consider $d_0 = n\kappa_0/N$. In the following we refer to this definition as model B, and to the former with $d_0=n/N$ as model A.
We show the corresponding phase transition curves in panels $(b)$ and $(d)$ of figure $\ref{fig:fig2}$.
Interestingly, even though with model B the number of links rewired at each time depends linearly on $\kappa_0$, our main results hold as with model B there is also an optimal intermediate $\kappa_0$ that requires a minimum time $\Delta$ to achieve a stationary memory state, indicating a non-linear effect of the initial density on the emerging behavior of the system.

\begin{figure}
\centering
\includegraphics[width=1.0\columnwidth]{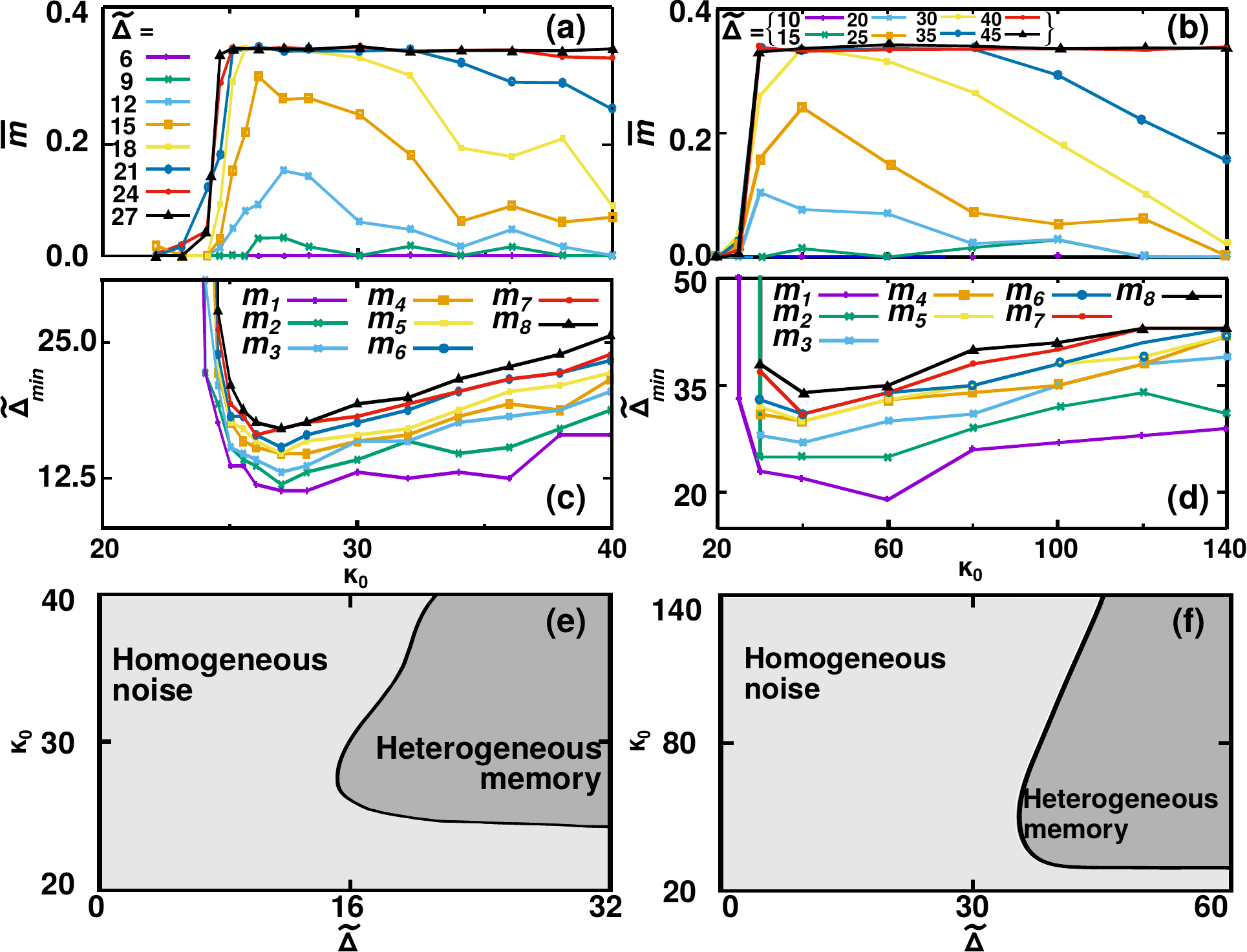}
\caption{Panels $(a)$ and $(b)$ show $\bar{m}(\kappa_0)$ for different values of $\Delta$, respectively for models A and B. Panels $(c)$ and $(d)$ show  $\widetilde{\Delta}^a_{\min}(\kappa_0)$ for different values of $\Delta$, respectively for models A and B.  
This shows how long the frozen period needs to be in order to obtain a certain mean stationary overlap, so that smaller times indicate a faster stabilization of the memory state. 
Panels $(e)$ and $(f)$ show the two dimensional phase diagram of the system, respectively for models A and B. These are  obtained through analysis of $\bar{m}(\widetilde{\Delta},\kappa_0)$ and  $\bar{g}(\widetilde{\Delta},\kappa_0)$, so that the heterogeneous memory phase is defined by  $\bar{m}>0$ (in particular we select $\bar{m} \geq 0.35$) and $\bar{g} \to 0$, whereas the homogeneous noisy one has $\bar{m}\to 0$ and $\bar{g}\to 1$. Parameter values as in figure $\ref{fig:fig2}$. $\widetilde{\Delta}$ for model B (panel $(f)$) is defined as $\widetilde{\Delta} = \Delta/1000$. 
\label{fig:fig3} }
\end{figure}

The non-linear (or second order) effect of $\kappa_0$ also implies that the complementary representation $\bar{m}(\kappa_0)$ for a given $\widetilde{\Delta}$ 
shows a bell shape for intermediate $\widetilde{\Delta}$, 
with a maximum at $\kappa_0^A\approx 27$ and $\kappa_0^B\approx 40$ respectively for models A and B (see figure $\ref{fig:fig3}$ panels $(a)$ and $(b)$). 
Similarly, one can explicitly measure the minimum value of $\widetilde{\Delta}$ needed to achieve memory for a given density. More generally, we define $\widetilde{\Delta}_{\min}^a(\kappa_0)$ as the minimum value of $\widetilde{\Delta}$ needed to reach a stationary mean overlap equal to $m_a = 0.1 a\ \bar{m}$, $a={1,2,..,10}$ for a given $\kappa_0$. 
This definition aims to measure how much time it takes for a given configuration to organize into the heterogeneous state. 
Minimal values of $\widetilde{\Delta}^a_{\min}$ indicate an optimal initial configuration to reach memory with a minimal energy consumption. 
Our measures (figure $\ref{fig:fig3}$ panels $(c)$ and $(d)$) indicate a minimum $\widetilde{\Delta}^a_{\min}$ for the same connectivity as the maximum of $\bar{m}(\kappa_0)$, namely $\kappa_0^A$ and $\kappa_0^B$ in both alternative model descriptions. 
Therefore, memory is not only reached faster but it is also stronger for the optimal $\kappa_0$.

Finally, an integrated view of the effect of network dynamics and the emergent behavior of the system can be obtained by the phase diagram of the system, as shown in figure $\ref{fig:fig4}e,f$, as a function of the control parameters $\widetilde{\Delta}$ and $\kappa_0$. 
These have been obtained by the integrated analysis of $\bar{m}(\widetilde{\Delta}, \kappa_0)$ and $\bar{g}(\widetilde{\Delta}, \kappa_0)$, which indicates the existence of a region of stationary memory and heterogeneous networks for both high $\widetilde{\Delta}$ and high $\kappa_0$. 
The phase transition between noise and memory moves to higher $\widetilde{\Delta}$ for higher $\kappa_0$ in an approximately linear manner, as previously discussed, leading to the contraction of the heterogeneous memory region. On the other hand, very small $\kappa_0$ never leads to memory, due to the high thermal noise.

Summing up, our results show the benefit of intermediate densities with respect to very high ones in order to achieve memory in a noisy environment. 
Interestingly, results hold when the density of synaptic turnover $d_0$ is re-scaled linearly with the density of synapses $\kappa_0$, indicating that the longer transient time needed to reach the heterogeneous memory phase for higher $\kappa_0$ is not only due to the higher number of synapses that need to be rewired. 
This result suggests why for an evolving network such as the infant brain it is detrimental to initially grow a very high density of synapses, since this increases the energy costs during growth and also during the pruning process, and it does not improve memory retrieval or network structure. On the contrary, a neural network with intermediate values of transient synaptic density would perform more efficiently during pruning. Moreover, this would also be  convenient in terms of energy consumption.

\subsection*{Transient heterogeneity determines network performance.}
We have shown a quadratic dependence of the stationary state on $\kappa_0$ (see figure $\ref{fig:fig3}$), and discussed the presence of multistability for intermediate values. However, what determines, on a given trial, the stationary state of the system?
Based on the results shown above, we propose that it is the transient level of heterogeneity (that is, the heterogeneity at the onset of the pruning) which determines the probability that the network will maintain memory. 

\begin{figure}
\centering
\includegraphics[scale=0.74]{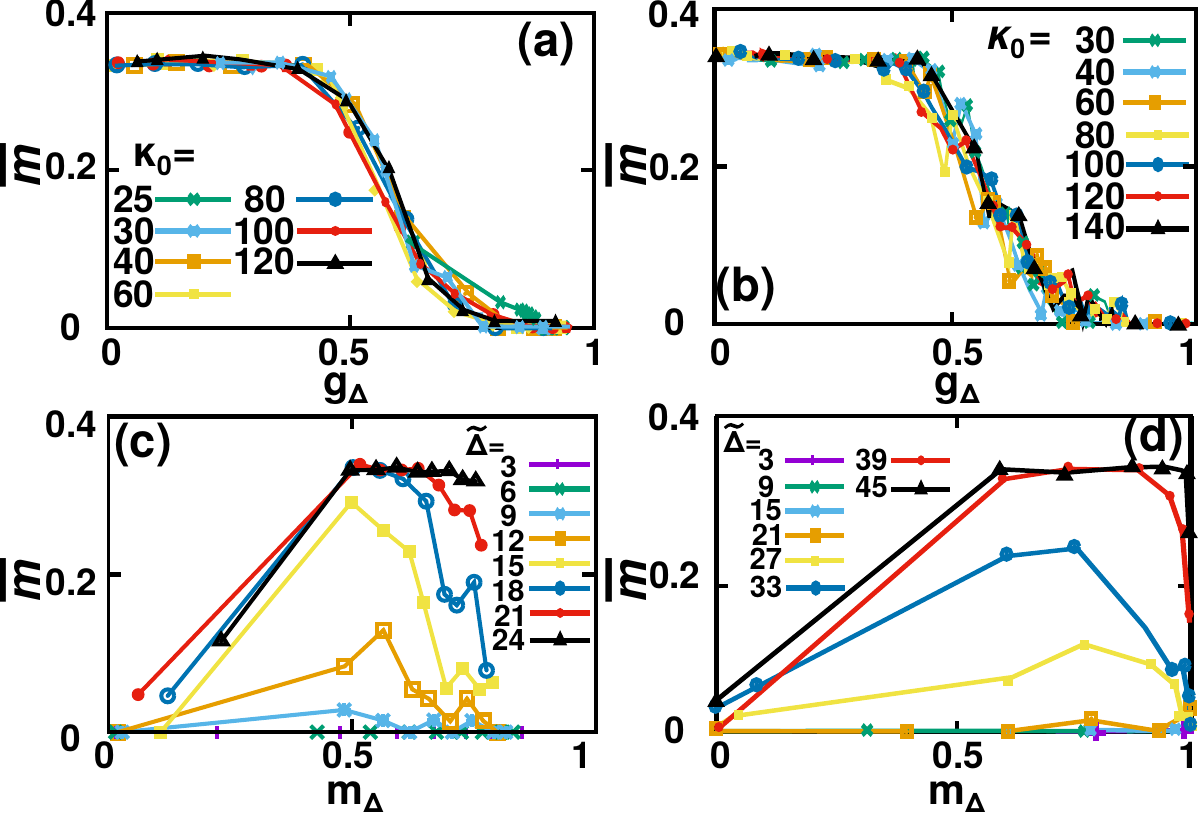}
\caption{ Panels (a) and (b) show $\bar{m}(g_\Delta(\widetilde{\Delta}))$ for different values of $\kappa_0$, respectively for models A and B. Panels (c) and (d) show $\bar{m}(m_\Delta(\kappa_0))$ for different values of $\widetilde{\Delta}$, respectively for models A and B. 
\label{fig:fig4} }
 
\end{figure}

In order to explore this hypothesis, we first define $g_{\Delta}$ and $m_{\Delta}$ as the values of $g$ and $m$ at the beginning of the pruning, that is, $g_{\Delta}=g(t=\Delta)$ and $m_\Delta = m(t=\Delta)$. 
These definitions allow us to explore how the stationary state depends on the transient evolution of the system. In particular, $\bar{m}(g_\Delta(\widetilde{\Delta}))$ (figure $\ref{fig:fig4}a,b$ respectively for models A and B) shows two main findings. Firstly,  a continuous transition from the heterogeneous memory state to the homogeneous noisy one as a function of $g_{\Delta}$; and, secondly, a collapse of the curves for different $\kappa_0$. 
In consequence, $g_\Delta$ determines whether the network will be able to maintain memory once the pruning begins: high onset heterogeneity (small $g_\Delta$) implies stationary memory, whereas low heterogeneity (high $g_\Delta$) implies a stationary noisy state. 
These results are independent of $\kappa_0$ and of the model used, indicating that $g_{\Delta}$ is a strong indicator of stationary memory. 

Notice that $g_\Delta$ depends not only on $\kappa_0$ and $\widetilde{\Delta}$, but also on $m_\Delta$, since the rewiring process only promotes heterogeneity when the network displays memory. 
This is shown in figure $\ref{fig:fig4}c,d$, where we display $\bar{m}\left( m_\Delta (\kappa_0) \right)$, with each curve corresponding to a given value of $\widetilde{\Delta}$, and respectively for models A and B. 
Since $m_\Delta$ does not unequivocally determine the stationary state, the curves do not collapse in this case. What it is obtained is a quadratic dependence of $\bar{m}$ on $m_\Delta$, indicating an optimal value of transient memory of $m_{\Delta}\approx 0.5$. 
This emerges because $m_\Delta\approx 0.5$ strongly correlates with a minimum $g_\Delta$, 
so that the value of $\widetilde{\Delta}_{\min}^a$ necessary for a stationary memory state is minimal for this value.
In this sense, if $m_{\Delta}\ll 0.5$, there is no transient memory because $\kappa_0$ is too small and, therefore, there is insufficient heterogeneity. On the other hand, if $m_{\Delta} \approx 1$, this is so because $\kappa_0$ is large and the network is still very homogeneous when pruning begins. In both cases, the network evolves towards a homogeneous configuration.
Note that this qualitative result does not depend on the particular definition of $d_0$ since it holds for both models A and B.


\subsection*{Discussion}
We present here an adaptive network model inspired by observed synaptic pruning that creates a dependence of the final network structure and performance on the transient synaptic density, as one should probably expect in nature. 
In this model, the introduction of a high fixed-density transient allows for network heterogenization and for the settlement of a memory state under 
noisy conditions.
We have analyzed in detail a point of the $(T,\alpha)$ phase space corresponding to a bistablity area between heterogeneous memory and homogeneous noisy states, where the system is most sensitive to the details of the evolution and the initial conditions. This happens to correspond to high noise $T$ and also to large $\alpha$, so that the system can heterogenize.

In these conditions, we have found that the model exhibits a discontinuous (first order) phase transition as one varies the length of the transient period of fixed density, $\widetilde{\Delta}$, and the value of this density, $\kappa_0$.
In fact, there is a quadratic effect of $\kappa_0$, such that medium values provide a faster and more stable evolution towards a memory stationary state and there is an optimal $\kappa_0$ that  optimizes the evolution into such a state. 
Therefore, our results could explain why real world networks such as those in the brain do not create enormous numbers of synapses to begin with during early development.
Moreover, being able to achieve eventual good performance with a limited density would also be preferable in terms of energy consumption. 
We have also shown that the transient heterogeneity determines the stationary state of the system. Given the aforementioned feedback loop, this depends on the transient memory $m_\Delta$ and $\kappa_0$, so that the stationary state of the system is ultimately determined by its physiological history.

We have also analyzed the robustness of the results to details of the model and the validity of the approximations considered. In particular, we have found a more realistic approximation such that the density of synaptic growth and death during the transient period of fixed connectivity is proportional to this connectivity. 
We have found that the qualitative behavior of the system does not depend on these details.
Namely, there is also a discontinuous transition from a homogeneous non-memory state to a heterogeneous memory one as the duration $\Delta$ of the transient period increases, whereas there is a non-linear dependence on $\kappa_0$, and the stationary state of the system is ultimately determined by $g_\Delta$. 
Therefore, the results obtained in this work -- namely the benefits of a transient high connectivity period, the second order effect of $\kappa_0$ and the crucial effect of $g_\Delta$ on the stationary state -- are robust with respect to microscopic details of the model.

A question this work clarifies is why brain development produces an initial growth of a great many synapses which are then gradually pruned. If the final density is optimal for energy consumption, why should one go through a transient state of twice this density? 
Our neural network with an evolving structure based on some simple biological considerations shows that the memory performance of the system does indeed depend on whether it passed through a transient period of relatively high synaptic density. 
A feedback loop thus emerges between neural activity and network topology such that, beginning with a random network, a transient state of high density can allow for the subsequent pruning of synapses as the topology is optimized for memory performance. Why, though, should the brain not begin with both the low density and high heterogeneity needed for good memory performance? 
We conjecture that much less genetic information is needed to build a random neural network that is subsequently shaped by its dynamics (under the influence of actual environmental conditions), than to specify a particular structure. 
This would also be a more robust developmental path. And why not begin with an even higher initial density? Apart from energetic considerations, we have also seen that there is an optimal initial synaptic density for a high performance neural network to emerge.

Our work here may serve as a starting point and a very suitable theoretical framework for studying the relationship that may exist between certain neurological disorders that appear during brain development, such as childhood autism and schizophrenia in young adults, and different synaptic pruning profiles, as has recently been suggested \cite{autismo1, schizo,schizo1}.

\subsection*{Acknowledgments}
We are grateful for financial support from the Spanish Ministry of Science and Technology and the ``Agencia Espa{\~n}ola de
Investigaci{\'o}n (AEI)''  under grant FIS2017-84256-P (FEDER funds), and from ``Obra Social La Caixa'' (ID 100010434 with code LCF/BQ/ES15/10360004).


\bibliographystyle{unsrt}

\newcounter{methods}
\renewcommand{\themethods}{Methods}

\subsection*[Methods]{Methods} 
\refstepcounter{methods}
\label{methods}

\textbf{The neural network model.}
Our system consists of an undirected $N$-node network whose edges change in discrete time. At time $t$, the adjacency matrix is $\lbrace e_{ij}(t)\rbrace$, $i,j=1,\ldots,N$, with elements $1$ or $0$ according to whether there exists or not an edge between the pair of nodes $(i,j)$, respectively.
The degree of node $i$ at time $t$ is $k_i(t)=\sum_j^N e_{ij}(t)$, and the mean degree of the network is $\kappa(t)=N^{-1}\sum_i^N k_i(t)$.

The neurons (nodes) are modeled as stochastic binary units, $s_i(t)=\{0,1\}$, indicating a firing or silent neuron, respectively.
Each edge $(i,j)$ is characterized by its synaptic weight $w_{ij}$, and the local field at neuron $i$ is $h_{i}(t)=\sum_{j=1}^{N}w_{ij}e_{ij}(t)s_{j}(t).$
The states of all neurons are updated in parallel at every time step following the Amari-Hopfield scheme, according to the transition probability 
$P\left\{ s_{i}(t+1)=1\right\} =\frac{1}{2}\left[1+\tanh\left(\beta\left[h_{i}(t)-\theta_{i}(t) \right]\right)\right],$
where $\theta_{i}(t)=\frac{1}{2}\sum_{j=1}^{N}w_{ij}e_{ij}(t)$ is a neuron's threshold for firing, and $\beta=T^{-1}$ is a noise parameter controlling stochasticity (analogous to the inverse temperature in statistical physics) \cite{amit1992modeling,bortz1975new}.

A  pattern  of activity  or memory $\left\{ \xi_{i}\right\}$, with mean $a_0=\left\langle\xi_i\right\rangle$, is encoded through $w_{ij}$ via the Hebbian learning rule, 
$w_{ij}=\left[\kappa_{\infty}a_{0}(1-a_{0})\right]^{-1} (\xi_{i} -a_{0})(\xi_{j} -a_{0}),$
where $\kappa_\infty=\kappa(t\rightarrow \infty)$.
The macroscopic order parameter is the overlap of the state of the network with   the memorized pattern, given by eq. $(\ref{eq:ov})$.
The canonical Amari-Hopfield model, which is here a reference, is defined on a fully connected network ($e_{ij}=1,\ \forall i\!=j$) and  exhibits a continuous phase transition at the critical value $T_c=1$ \cite{amit1992modeling}.

\textbf{The pruning model.}
Edge dynamics are modeled as follows. At each time $t$, each node has a probability $P_{i}^{g}=u\left(\kappa\right)\pi\left(I_{i}\right)$ of being assigned a new edge to another node, randomly chosen. Likewise, each node has a probability $P_{i}^{l}=d\left(\kappa\right)\eta\left(I_{i}\right)$ of losing one of its edges, randomly chosen. Here the time dependence has been dropped for clarity, and $I_i = |h_i - \theta_i|$ is a physiological variable that characterizes the local dependence.

Initially, network topology is defined as in the configuration model \cite{configurationmodel1995}. 
Time evolution is then accomplished in practice via computer simulations as follows.
First, the number of links to be created and destroyed is chosen according to two Poisson distributions with means $Nu\left(\kappa\right)$ and $Nd\left(\kappa\right)$, respectively. 
Then, as many times as needed according to this draw, we choose a node $i$ with probability $\pi\left(I_{i}\right)$ to be assigned a new edge, to another node randomly chosen;
and similarly we choose a new node $j$ according to $\eta\left(I_{j}\right)$ to lose an edge from one of its neighbors, randomly chosen.
This procedure uses the BKL algorithm to assure proper evolution towards stationarity \cite{bortz1975new}. 
Therefore, each node can then gain (or lose) an edge via two paths, and we define the effective values of $\pi$ and $\eta$ to account for this effect;
$\widetilde{\pi} = 1/2 \left( \pi\left(I_{i}\right)+1/N \right)$
and $\widetilde{\eta} = 1/2 \left (\eta\left(I_{i}\right)+k_{i}/(\kappa N) \right)$, where the $1/2$ factor is included to assure normalization. 

For the sake of simplicity, we  consider $\widetilde{\pi}$ and $\widetilde{\eta}$ to be power-law distributed \cite{johnson2010evolving}, which allows one to move smoothly from a sub-linear to a super-linear dependence with a single parameter, $\widetilde{\pi} =  I_{i}^{\alpha} / (\left\langle I^{\alpha} \right\rangle N)$ and $\widetilde{\eta} =  I_{i}^{\gamma} / (\left\langle I^{\gamma} \right\rangle N)$. Previous studies have shown that the qualitative results only depend on the ratio between $\gamma$ and $\alpha$, so that for simplicity we fix $\gamma=1$, and leave $\alpha$ as a topological control parameter.

\textbf{Time evolution of the mean connectivity.}
The time evolution of $\kappa(t)$ can be obtained in the topological limit in the model, defined by making the substitution $I_{i}\rightarrow k_{i}$, so that $\widetilde \eta_{i}=\widetilde \eta\left(k_{i}\right)$ and $\widetilde \pi_{i}=\widetilde\pi\left(k_{i}\right)$. 
This is possible since in the memory regime  $I_i \propto k_i$, whereas in the noisy one both variables follow equivalent noisy dynamics \cite{yo2017,FCN}.
In this way one can construct a master equation for $p(k,t)$ by considering network evolution as a one step process with transition rates $u(\kappa)\widetilde\pi(k)$ for degree increment and $d(\kappa)\widetilde\sigma(k)$ for the decrement. Approximating the temporal derivative for the expected value of the difference in a given $p(k,t)$ at each time step we get:
$
\frac{dp(k,t)}{dt} =  
\ u\left(\kappa\right)\widetilde{\pi}\left(k-1\right)p\left(k-1,t\right) 
+d\left(\kappa\right)\widetilde{\eta}\left(k+1\right)p\left(k+1,t\right) 
 -\left[u\left(\kappa\right)\widetilde{\pi}\left(k\right) +d\left(\kappa\right)\widetilde{\eta}\left(k\right)\right]p\left(k,t\right), 
$
which is exact in the limit of no degree-degree correlations between nodes.
From here it follows that $d\kappa(t) / dt = 2 \left[ u\left(\kappa(t)\right) - d\left( \kappa(t)\right) \right]$.
The time scale for structure changes is set by the parameter $n$, whereas the time unit for activity changes, $h_s$, is the number of Monte Carlo Steps (MCS) that the states of all neurons are updated according to the Hopfield dynamics between each structural network update.
Previous studies show a low dependence on this parameter in the cases of interest, so we report results here for $h_{s}=10$ MCS \cite{yo2017,FCN}.

\textbf{Statistics and general methods} In this work, we used systems sizes $N=800$, $1600$ and $3200$, as indicated in each section. Results for $N<800$ presented strong finite size effects, so they were discarded (data not shown). The sample size for each result was chosen by convergence of the mean value. 
Measures of the global variables on the stationary state are obtained by averaging during a long window of time: $\bar{f} = \Delta t^{-1} \sum_{t=t_0}^{t_0+\Delta t} f(t)$.

\textbf{Code availability} Generated codes are available from the corresponding author upon reasonable request.

\textbf{Data availability} All data that support this study are available from the corresponding author upon reasonable request.

\textbf{Author contributions} A.P.M, J.J.T., S.J. and J.M designed the model and the analyses. A.P.M. wrote the code and performed the analysis.

\textbf{Competing interests} The authors declare no competing interests.

\end{document}